\documentclass[aps,prb,a4paper,preprint,showpacs]{revtex4}
\usepackage{amsmath}
\usepackage{amsfonts}
\usepackage{amssymb}
\usepackage{slashbox}
\usepackage{graphicx}
\usepackage{amsbsy}

\begin{document}
\title{Fano Resonance in the Nonadiabatically Pumped Shot Noise of a Time-Dependent Quantum Well in 2DEG and Graphene }
\author{Rui Zhu$^1$\renewcommand{\thefootnote}{*}\footnote{Corresponding author.
Electronic address:
rzhu@scut.edu.cn}, Jiao-Hua Dai$^1$, and Yong Guo$^{2,3}$}
\address{$^1$Department of Physics, South China University of Technology,
Guangzhou 510641, P. R. China \\
$^2$Department of Physics and State Key Laboratory of Low-Dimensional Quantum Physics, Tsinghua University, Beijing
100084, P. R. China \\ $^3$ Collaborative Innovation Center of Quantum Matter, Beijing, P. R. China}

\begin{abstract}

Interference between different quantum paths can generate Fano resonance. One of the examples is transport through a quasibound state driven by time-dependent scattering potential. Previously it is found that Fano resonance occurs as a result of energy matching in one-dimensional systems. In this work, we demonstrate that when transverse motion is present, Fano resonance occurs precisely at the wavevector matching situation. Using the Floquet scattering theory, we considered the transport properties of a nonadiabatic time-dependent well both in the 2DEG and monolayer graphene structure. Dispersion of the quasibound state of a static quantum well is obtained with transverse motion present. We found that Fano resonance occurs when the wavevector in the transport direction of one of the Floquet sidebands is exactly identical to that of the quasibound state in the well at equilibrium and follows the dispersion pattern of the latter. To observe the Fano resonance phenomenon in the transmission spectrum, we also considered the pumped shot noise properties when time and spatial symmetry secures vanishing current in the considered configuration. Prominent Fano resonance is found in the differential pumped shot noise to the reservoir Fermi energy.

\end{abstract}


\maketitle

\narrowtext

\section{Introduction}

Fano resonance is a widely present phenomenon in atomic spectrum, light propagation, quantum transport, matter-wave scattering in ultracold atom
systems, and etc\cite{MiroshnichenkoRMP2010, FanoPR1961}. It can be interpreted by destructive interference of transport in different quantum paths especially when a discrete state interacts with a continuum of propagating modes. In quantum transport systems, when a donor impurity is embedded in a free conductor, Fano resonance can result from path interference, with additional quantum paths necessary for interference generated by spatially or time dependent potential. Floquet sidebands are formed in nonadiabatic quantum pumping driven by one or several high-frequency oscillating potentials. In the case of a time-dependent quantum well, when one of the Floquet levels matches the quasibound level inside there strikes a Fano resonance\cite{LiPRB1999, DaiEPJB2014, WTLuJAP2012}, which can be observed in the pumped shot noise\cite{DaiEPJB2014}. These previous work proposes that the Fano resonance occurs as a result of energy level matching between one of the Floquet sidebands and one of the quasibound states within the well in one-dimensional transport. It is unclear whether the Fano resonance occurs with energy matching or wavevector matching or other mechanisms when transverse motion enters. Earlier Fano resonance in the momentum space was already discussed in topological filters\cite{MiroshnichenkoRMP2010, BurioniPRE2006} and quadratic waveguide arrays\cite{MiroshnichenkoRMP2010, MiroshnikenkoOL2005}. In this work, we would investigate the Fano resonance properties in nonadiabatic quantum pumping driven by a single oscillating electric potential well in 2DEG (two-dimensional electron gas) and graphene with nonzero transverse wavevectors.

The two systems we would consider are 2DEG and monolayer graphene. 2DEG is a single-particle model of free electron states with parabolic energy-momentum dispersion. It can be formed in semiconductor heterostructures and is a general platform for various quantum phenomenons. The quasiparticle states in a monolayer Graphene sheet can also be modeled as a two-dimensional free gas\cite{KatsnelsonNP2006}. The difference is that it has ``light-cone"-like dispersion with the conduction and valence band connected at the Dirac point. As a result, hole states as well as electrons contribute to its transport properties. Its low energy behavior can be described by the Dirac equation.  From the band structure investigation of the monolayer graphene\cite{NakadaPRB1996}, its experimental realization\cite{NovoselovNature2005}, the quantum spin Hall effect\cite{KanePRL2005}, to the quantum anomalous Hall state in bilayer graphene\cite{NandkishorePRB2010}, graphene has aroused unceasing interest among physicists. As an important dynamic transport process, quantum pumping properties in graphene-based structures were also targeted from different view angles\cite{SanJosePRB2011, ZhuAPL2009, PradaPRB2009, ZhuJPCM2011}. Recently, irradiation induced Floquet topological transitions in graphene also attracted attention\cite{UsajArxiv2014, DelplacePRB2013}. With the Dirac fermions being the charge-carrying quasiparticles, the graphene has its unique significance in transport investigations.

Shot noise measures the current fluctuations originated from quantization of charge carriers\cite{BlanterPR2000}. In the past two decades, shot noise has played an important role in charge confirmation such as of Cooper pairs\cite{LeflochPRL2003} and the Laughlin quasiparticles\cite{SaminadayarPRL1997}. Similar to bias driven conductance, the parametric pumped charge current is also accompanied by the pumped shot noise featuring physical information beyond conductance measurements\cite{ZhuJPCM2011, MoskaletsPRB2004, ZhuPRB2010, DaiEPJB2014}. The Klein paradox\cite{ZhuJPCM2011}, Fano resonance\cite{DaiEPJB2014}, and lifetime of the quasistationary states between magnetic domain walls\cite{ZhuPRB2010} were found to be characterized in the shot noise, which is complementary or beyond the conductance properties.

Although intensive theoretical work has been done on the bias driven shot noise properties of various mesoscopic conductors\cite{ZhuNova2011} and the general scattering theory for adiabatic\cite{MoskaletsPRB2002} and nonadiabatic pumped shot noise\cite{MoskaletsPRB2004} is derived, the specific pumped shot noise properties in different quantum transport systems are less covered. They represent the underlying physics of different materials and devices, some of which is beyond conductance information. In the pumping process driven by time dependent external or internal parameters, virtual or temporary transmission within a cyclic period generates considerable noise even when time and spatial reversal symmetry secures vanishing time-averaged charge current. The Floquet scattering theory was already used to investigate quantum pumping behavior in the 2DEG\cite{LiPRB1999, SanJosePRB2011} and graphene\cite{SanJosePRB2011} structures. Recently, the Floquet-Bloch theory\cite{UsajArxiv2014, DelplacePRB2013} and the Floquet-Magnus approach\cite{LopezPRB2012} were developed respectively to investigate radiation induced band topology manipulation and the eigenstates modulation under ac-driven spin-orbit interaction both in monolayer graphene. However, neither of them considered the transport properties. In this work, we follow Li \emph{et al}.'s Floquet transmission\cite{LiPRB1999} and Moskalets \emph{et al}.'s Floquet shot noise\cite{MoskaletsPRB2004} frameworks and extend them to two-dimensional free electron gases and Dirac fermions of graphene. Fano resonance in the nonadiabatically pumped shot noise at quasibound wavevectors driven by an ac single-well potential was found with no dc charge current pumped out due to time and spatial reversal symmetry.

Other parts of the paper would be organized as follows. Discussions of the 2DEG and graphene would be given in Section II and III, respectively. Analysis of the confined states, Fano resonance properties of the transmission under the Floquet theory, and the pumped shot noise properties would be three subsections of them. A conclusion would be given in Section IV.

\section{Nonadiabatic quantum pumping in 2DEG}

We consider the nonadiabatic pumping properties in a 2DEG driven by a time-dependent electric potential well $V(t)=-V_0+V_1 \cos (\omega t)$ with width $L$. $V_0$ is the static well depth, $V_1$ is the driving amplitude, and $\omega$ is the driving frequency. The considered device is sketched in Fig. 1. Assuming the 2DEG located in the $x$-$y$ plane, the time-dependent Hamiltonian of the electrons can be expressed as:
\begin{equation}
H\left( t \right) =  - \frac{{{\hbar ^2}}}{{2{m^*}}}\left( {\frac{{{\partial ^2}}}{{\partial {x^2}}} + \frac{{{\partial ^2}}}{{\partial {y^2}}}} \right) + U\left( {x,t} \right),
\end{equation}
with
\begin{equation}
U\left( {x,t} \right) = \left\{ {\begin{array}{*{20}{l}}
{V\left( t \right),}&{0 \le x \le L,}\\
{0,}&{{\rm{others}}.}
\end{array}} \right.
\label{Uxt}
\end{equation}
For GaAs the electron effective mass $m^*$=0.067$m_e$, where $m_e$ is the mass of the free electron and our
discussion is based on single electron approximation and
coherent tunneling.

\subsection{Quasibound States within a Static Quantum Well}

In advance of the time-dependent treatment, we consider the quasibound states within the static quantum well with width $L$ and depth $V_0$ spanned in the $x$ direction in the 2DEG. We set the energy coordinate to be $- V_0 $ at the bottom of the well. When the electron
is confined in the well with its energy $E>-V_0$, the wave functions inside and outside of the well can be written as
\begin{equation}
\psi \left( {x,y} \right) = {e^{i{k_y}y}}\left\{ {\begin{array}{*{20}{l}}
{r{e^{\kappa x}},}&{x \le 0,}\\
{a{e^{i{k_x}x}} + b{e^{ - i{k_x}x}},}&{0 \le x \le L,}\\
{t{e^{ - \kappa x}},}&{x \ge L,}
\end{array}} \right.
\end{equation}
where ${k_x} = {{\sqrt {2{m^*}\left( {E + {V_0}} \right) - {\hbar ^2}k_y^2} } \mathord{\left/
 {\vphantom {{\sqrt {2{m^*}\left( {E + {V_0}} \right) - {\hbar ^2}k_y^2} } \hbar }} \right.
 \kern-\nulldelimiterspace} \hbar }$ and $\kappa  = {{\sqrt { - 2{m^*}E - {\hbar ^2}k_y^2} } \mathord{\left/
 {\vphantom {{\sqrt { - 2{m^*}E - {\hbar ^2}k_y^2} } \hbar }} \right.
 \kern-\nulldelimiterspace} \hbar }$. Continuity equations of the wave functions and their derivatives at $x=0$ and $x=L$ are:
\begin{equation}
 \left\{ \begin{array}{l}
r = a + b,\\
\kappa r = i{k_x}a - i{k_x}b,\\
a{e^{i{k_x}L}} + b{e^{ - i{k_x}L}} = t{e^{ - \kappa L}},\\
i{k_x}a{e^{i{k_x}L}} - i{k_x}b{e^{ - i{k_x}L}} =  - \kappa t{e^{ - \kappa L}}.
\end{array} \right.
\end{equation}
Solvability of these equations gives rise to the secular
equation
\begin{equation}
\xi  = \left| {\begin{array}{*{20}{c}}
1&{ - 1}&{ - 1}&0\\
\kappa &{ - i{k_x}}&{i{k_x}}&0\\
0&{{e^{i{k_x}L}}}&{{e^{ - i{k_x}L}}}&{ - {e^{ - \kappa L}}}\\
0&{i{k_x}{e^{i{k_x}L}}}&{ - i{k_x}{e^{ - i{k_x}L}}}&{\kappa {e^{ - \kappa L}}}
\end{array}} \right| = 0.
\end{equation}
Roots of $E$ for this equation are the quasibound state energies.
They can be obtained numerically by the sign-reversal points of ${{\partial \left| \xi  \right|} \mathord{\left/
 {\vphantom {{\partial \left| \xi  \right|} {\partial E}}} \right.
 \kern-\nulldelimiterspace} {\partial E}}$. The quasibound
levels as a function of $k_y$ is shown in Fig. 2. There are two quasibound levels within the well. We label the energy of the quasibound state as $E_b$. Decreasing parabolic dispersion can be seen in the high quasibound level. As $k_y$ increases, transverse motion costs larger energy giving rise to the decrease in $E_b$. Parabolic dispersion pattern is natural as a result of parabolic conduction band of 2DEG. Both quasibound levels vanishes when the wave vector in the transport direction $k_x$ becomes imaginary.

\subsection{Floquet Scattering}

We use the Floquet scattering theory to investigate the nonadiabatic
quantum pump driven by the time-dependent
well potential\cite{LiPRB1999, DaiEPJB2014}. Wave functions in the three scattering regions can be written as:
\begin{equation}
\psi \left( {x,y,t} \right) = {e^{i{k_y}y}}\sum\limits_{n =  - \infty }^{ + \infty } {{e^{{{ - i{E_n}t} \mathord{\left/
 {\vphantom {{ - i{E_n}t} \hbar }} \right.
 \kern-\nulldelimiterspace} \hbar }}}\left\{ {\begin{array}{*{20}{l}}
{a_n^l{e^{i{k_{xn}}x}} + b_n^l{e^{ - i{k_{xn}}x}},}&{x \le 0,}\\
{\sum\limits_{m =  - \infty }^{ + \infty } \begin{array}{l}
\left( {{a_m}{e^{i{\kappa _m}x}} + {b_m}{e^{ - i{\kappa _m}x}}} \right)\\
 \times {J_{n - m}}\left( {\frac{{V_1}}{{\hbar \omega }}} \right),
\end{array} }&{0 \le x \le L,}\\
{a_n^r{e^{ - i{k_{xn}}x}} + b_n^r{e^{i{k_{xn}}x}},}&{x \ge L.}
\end{array}} \right.}
\end{equation}
The potential is translation invariant in the $y$-direction. Plane wave with $k_y$ preserved can be assumed during transmission. The incident and outgoing
electron waves consist of infinite Floquet sidebands,
as shown in Fig. 1. These sidebands are formed by motion in the $x$ and $y$ directions. The Floquet state energies are $E_n =E_F +n \hbar \omega$. $E_F$ is the Fermi energy of the left and right electrodes at the two sides of the oscillating well with no bias between them. The sideband index $n$ is an integer varying from $- \infty$ to $+ \infty$ in an ideal
exactness. Numerical accuracy is secured for its cutoff\cite{LiPRB1999} $N> V_1 / (\hbar \omega)$. In this case we set $N=5$. The Floquet wave vectors ${k_{xn}} = {{\sqrt {2{m^*}{E_n} - {\hbar ^2}k_y^2} } \mathord{\left/
 {\vphantom {{\sqrt {2{m^*}{E_n} - {\hbar ^2}k_y^2} } \hbar }} \right.
 \kern-\nulldelimiterspace} \hbar }$ and ${\kappa _m} = {{\sqrt {2{m^*}\left( {{E_m} + {V_0}} \right) - {\hbar ^2}k_y^2} } \mathord{\left/
 {\vphantom {{\sqrt {2{m^*}\left( {{E_m} + {V_0}} \right) - {\hbar ^2}k_y^2} } \hbar }} \right.
 \kern-\nulldelimiterspace} \hbar }$. $J_n (x)$ are the $n$-th order first kind Bessel functions.
Here, different from the one-dimensional case, $k_{xn}$ is imaginary meaning an
evanescent mode even when $E_n >0$ if $k_y$ is relatively large. Transmission for this channel vanishes. $a_n^{l/r}$ and $b_n^{l/r}$ are the probability amplitudes of waves flowing out of and into the left/right
electrodes, respectively.

The Floquet
scattering matrix ${s_{\alpha \beta }}\left( {{E_n},{E_m}} \right)$ can be obtained\cite{LiPRB1999, DaiEPJB2014} by continuity of $\psi $ and ${{\partial \psi } \mathord{\left/
 {\vphantom {{\partial \psi } {\partial x}}} \right.
 \kern-\nulldelimiterspace} {\partial x}}$ at the boundaries of the oscillating quantum well $x=0$ and $L$. It connects annihilation operators ${\hat a_\alpha }\left( E \right)$ and ${\hat b_\alpha }\left( E \right)$ of the incident
and outgoing electrons to the driven potential as
\begin{equation}
{\hat b_\alpha }\left( {{E_n}} \right) = \sum\limits_{m,\beta } {{s_{\alpha \beta }}\left( {{E_n},{E_m}} \right){{\hat a}_\beta }\left( {{E_m}} \right)} .
\end{equation}
The total Floquet transmission probability follows as
\begin{equation}
{T_F} = \sum\limits_{n = 0}^N {{{\left| {{s_{RL}}\left( {{E_F},{E_n}} \right)} \right|}^2}}.
\end{equation}
Under real parameter settings, numerical results of $T_F$ were shown in panel (a) of Fig. 3. Sharp Fano resonance can be seen in the transmission spectrum as a function of $E_F$ when one of the Floquet channel matches the quasibound level confined in the well. For larger $k_y$, it occurs at higher Fermi energies. It could be understood as transverse motion energy is supplied by the total energy of the incident electron. To see relation between the Fano resonance and the quasibound state, the Fano resonance occurred Fermi energy $E_{\rm{Fano}}$ as a function of $k_y$ is plotted in Fig. 2. Parabolic dispersion pattern is obvious. It follows the relation
\begin{equation}
{E_{{\rm{Fano}}}} - \hbar \omega  - \frac{{{\hbar ^2}k_y^2}}{{{m^*}}} = {E_b}.
\label{FanoEb}
\end{equation}
Re-obtained $E_b$ from $E_{\rm{Fano}}$ by Eq. (\ref{FanoEb}) is shown by blue asterisks in Fig. 2. Within numerical accuracy, the two data are identical. It can be seen that by external driving potential the incident electron emits an energy quantum of $\hbar \omega$, releases the transverse motion energy ${{{\hbar ^2}k_y^2} \mathord{\left/
 {\vphantom {{{\hbar ^2}k_y^2} {2{m^*}}}} \right.
 \kern-\nulldelimiterspace} {2{m^*}}}$, enters the quasibound state, supplies the transverse motion energy for the bound state ${{{\hbar ^2}k_y^2} \mathord{\left/
 {\vphantom {{{\hbar ^2}k_y^2} {2{m^*}}}} \right.
 \kern-\nulldelimiterspace} {2{m^*}}}$, and bounces back to the $E_F$ channel. This path interferes with direct tunneling giving rise to a Fano resonance. Therefore, in nonadiabatic quantum pumping of 2DEG with transverse motion present, the Fano resonance occurs as a result of transport wave vector matching.

\subsection{Pumped Shot Noise}

In our potential configuration of a single time-dependent quantum well, spatial and time-reversal symmetry secures zero pumped current at no electric or temperature bias between the left and right electrodes. However, the shot noise measuring the current fluctuation can be considerably large due to virtual transport of electrons and holes during one driving cycle\cite{ZhuJPCM2011, ZhuPRB2010, DaiEPJB2014}.
With the Floquet scattering matrix obtained, the zero-frequency nonadiabatic pumped shot noise measuring current fluctuation correlation between particle beams from $\alpha$ and $\beta$ electrodes can be expressed as\cite{MoskaletsPRB2004}
\begin{equation}
{S_{\alpha \beta }} = \frac{{{e^2}}}{h}\int_0^\infty  {dE\sum\limits_{\gamma \delta } {\sum\limits_{m,n,p =  - \infty }^{ + \infty } {{M_{\alpha \beta \gamma \delta }}\left( {E,{E_m},{E_n},{E_p}} \right){{\left[ {{f_0}\left( {{E_n}} \right) - {f_0}\left( {{E_m}} \right)} \right]}^2}} } } ,
\label{Noise}
\end{equation}
with
\begin{equation}
{M_{\alpha \beta \gamma \delta }}\left( {E,{E_m},{E_n},{E_p}} \right) = s_{\alpha \gamma }^*\left( {E,{E_n}} \right){s_{\alpha \delta }}\left( {E,{E_m}} \right)s_{\beta \delta }^*\left( {{E_p},{E_m}} \right){s_{\beta \gamma }}\left( {{E_p},{E_n}} \right).
\end{equation}
As a result of particle flux conservation, the Floquet scattering matrix is unitary\cite{DaiEPJB2014} and the pumped shot noise has the symmetry of
${S_{LL}} =  - {S_{LR}} =  - {S_{RL}} = {S_{RR}}.$ Our numerical treatment considers one of them and label $S \equiv {S_{LL}}$.

Variation of the pumped shot noise as a function of $E_F$ for different $k_y$ is shown in panel (b) of Fig. 3. The shot noise increases with the Fermi energy as a result of more energy channels contributing to the transport. For even larger Fermi energy, the active
Floquet bands are out of the potential well. Its influence becomes weak and transmission is nearly ballistic. The shot noise
decreases. For larger $k_y$, transverse motion consumes more incident energy, the shot noise curve as well as the total Floquet transmission translates along the Fermi energy axis. A slight inflection can be seen in the pumped shot noise at the Fano resonance occurring Fermi energy. The pumped shot noise is a result of transmission of all energy channels below the Fermi energy. Therefore, contribution from the Fano resonance channel is weak. However, if we
differentiate $S$ as a function of the Fermi energy and let ${S_d} = {{\partial S} \mathord{\left/
 {\vphantom {{\partial S} {\partial {E_F}}}} \right.
 \kern-\nulldelimiterspace} {\partial {E_F}}}$, sharp resonance reemerges exactly at the resonance energy of the transmission, which is shown in panel (c) of Fig. 3. By doing the differentiate, contribution by a single energy channel is visible. The shot noise reflects virtual transport processes within a driving cycle even when vanishing pumped current is secured by spatial and time-reversal symmetry in this single-driving-parameter case. Hence, the Fano resonance in the total Floquet transmission is recorded in the pumped shot noise spectrum. Especially the Fano resonance pattern reappears in the differential pumped shot noise.

\section{Nonadiabatic quantum pumping in monolayer graphene}

We consider the nonadiabatic quantum pumping properties driven by a time-dependent quantum well at zero bias in monolayer graphene. The potential profile is sketched in Fig. 1, which is identical to the previous consideration of 2DEG.
 For graphene, the time-dependent Dirac equation outside and within the oscillating potential well can be written as\cite{SavelevPRL2012}
\begin{equation}
i\hbar \frac{\partial }{{\partial t}}\psi \left( {x,y,t} \right) = \left[ {{v_F}\left( {{\bf{\sigma }} \cdot {\bf{\hat p}}} \right) + U\left( {x,t} \right)} \right]\psi \left( {x,y,t} \right),
\end{equation}
with the potential profile in space and time $U\left( {x,t} \right)$ identical to that of Eq. (\ref{Uxt}).
${v_F} \approx {10^6}$ $\rm{m s}^{-1}$ is the Fermi velocity and ${\bf{\sigma }} = \left( {{\sigma _x},{\sigma _y}} \right)$
are the Pauli matrices.

\subsection{Quasibound States within a Static Quantum Well}

In advance of the time-dependent treatment, we consider the quasibound states confined in the static quantum well with width $L$ and depth $V_0$ spanned in the $x$ direction of monolayer graphene. We set the energy coordinate to be $- V_0 $ at the bottom of the well. The electron-hole spinor states $\psi  = {e^{i{k_y}y}}\left( {{\psi _1},{\psi _2}} \right)'$ with energy $E > - V_0$ inside and outside the well can be written as
\begin{equation}
{\psi _1} = \left\{ {\begin{array}{*{20}{l}}
{r{e^{qx}},}&{x \le 0,}\\
{a{e^{i{k_x}x}} + b{e^{ - i{k_x}x}},}&{0 \le x \le L,}\\
{t{e^{ - qx}},}&{x \ge L,}
\end{array}} \right.
\end{equation}
\begin{equation}
{\psi _2} = \left\{ {\begin{array}{*{20}{l}}
{ - sr{e^{qx - i\theta }},}&{x \le 0,}\\
{s'\left( {a{e^{i{k_x}x + i\phi }} - b{e^{ - i{k_x}x - i\phi }}} \right),}&{0 \le x \le L,}\\
{st{e^{ - qx + i\theta }},}&{x \ge L,}
\end{array}} \right.
\end{equation}
with $s = {\mathop{\rm sign}\nolimits} \left( E \right)$, $s' = {\mathop{\rm sign}\nolimits} \left( {E + {V_0}} \right)$, ${k_x} = {{\sqrt {{{\left( {E + {V_0}} \right)}^2} - {{\left( {\hbar {v_F}{k_y}} \right)}^2}} } \mathord{\left/
 {\vphantom {{\sqrt {{{\left( {E + {V_0}} \right)}^2} - {{\left( {\hbar {v_F}{k_y}} \right)}^2}} } {\left( {\hbar {v_F}} \right)}}} \right.
 \kern-\nulldelimiterspace} {\left( {\hbar {v_F}} \right)}}$, $q = {{\sqrt {{E^2} - {{\left( {\hbar {v_F}{k_y}} \right)}^2}} } \mathord{\left/
 {\vphantom {{\sqrt {{E^2} - {{\left( {\hbar {v_F}{k_y}} \right)}^2}} } {\left( {\hbar {v_F}} \right)}}} \right.
 \kern-\nulldelimiterspace} {\left( {\hbar {v_F}} \right)}}$, $\theta  = {\tan ^{ - 1}}\left( {{{ - i{k_y}} \mathord{\left/
 {\vphantom {{ - i{k_y}} q}} \right.
 \kern-\nulldelimiterspace} q}} \right)$, and $\phi  = {\tan ^{ - 1}}\left( {{{{k_y}} \mathord{\left/
 {\vphantom {{{k_y}} {{k_x}}}} \right.
 \kern-\nulldelimiterspace} {{k_x}}}} \right)$.

 By continuity of the spinor wave function at the boundaries of the quantum well, secular equation for the eigenenergy can be written as
 \begin{equation}
 \xi  = \left| {\begin{array}{*{20}{c}}
1&{ - 1}&{ - 1}&0\\
{ - s{e^{ - i\theta }}}&{ - s'{e^{i\phi }}}&{s'{e^{ - i\phi }}}&0\\
0&{{e^{i{k_x}L}}}&{{e^{ - i{k_x}L}}}&{ - {e^{ - qL}}}\\
0&{s'{e^{i{k_x}L + i\phi }}}&{ - s'{e^{ - i{k_x}L - i\phi }}}&{ - s{e^{ - qL + i\theta }}}
\end{array}} \right| = 0.
\end{equation}
Numerical results of the secular equation is given in the small black dots in Fig. 4, which reproduces results of Ref. \onlinecite{PereiraPRB2006}.

\subsection{Floquet Scattering}

To investigate the influence of the quasibound states to the nonadiabatic quantum pumping properties, we try the Floquet Dirac spinor $\psi  = {e^{i{k_y}y}}\left( {{\psi _1},{\psi _2}} \right)'$ in the following form\cite{LiPRB1999, KatsnelsonNP2006}
\begin{equation}
{\psi _1}\left( {x,t} \right) = \sum\limits_{n =  - \infty }^{ + \infty } {{e^{ - {{i{E_n}t} \mathord{\left/
 {\vphantom {{i{E_n}t} \hbar }} \right.
 \kern-\nulldelimiterspace} \hbar }}}\left\{ {\begin{array}{*{20}{l}}
{A_n^i{e^{i{k_{xn}}x}} + A_n^o{e^{ - i{k_{xn}}x}},}&{x \le 0,}\\
{\sum\limits_{m =  - \infty }^{ + \infty } {\left( {{a_m}{e^{i{q_m}x}} + {b_m}{e^{ - i{q_m}x}}} \right){J_{n - m}}\left( {\frac{{{V_1}}}{{\hbar \omega }}} \right)} ,}&{0 \le x \le L,}\\
{B_n^i{e^{ - i{k_{xn}}x}} + B_n^o{e^{i{k_{xn}}x}},}&{x \ge L,}
\end{array}} \right.}
\label{WaveFunction1}
\end{equation}
\begin{equation}
{\psi _2}\left( {x,t} \right) = \sum\limits_{n =  - \infty }^{ + \infty } {{e^{ - {{i{E_n}t} \mathord{\left/
 {\vphantom {{i{E_n}t} \hbar }} \right.
 \kern-\nulldelimiterspace} \hbar }}}\left\{ {\begin{array}{*{20}{l}}
{A_n^i{s_n}{e^{i{k_{xn}}x + i{\phi _n}}} - A_n^o{s_n}{e^{ - i{k_{xn}}x - i{\phi _n}}},}&{x \le 0,}\\
{\sum\limits_{m =  - \infty }^{ + \infty } {\left( {s{'_m}{a_m}{e^{i{q_m}x + i{\theta _m}}} - s{'_m}{b_m}{e^{ - i{q_m}x - i{\theta _m}}}} \right){J_{n - m}}\left( {\frac{{{V_1}}}{{\hbar \omega }}} \right)} ,}&{0 \le x \le L,}\\
{ - B_n^i{s_n}{e^{ - i{k_{xn}}x - i{\phi _n}}} + B_n^o{s_n}{e^{i{k_{xn}}x + i{\phi _n}}},}&{x \ge L,}
\end{array}} \right.}
\label{WaveFunction2}
\end{equation}
which secures identical spinor normalization for all Floquet orders of a constant $\sqrt {2}$. In the infinite graphene layer, the potential is homogeneous in the $y$ direction, therefore a plane wave component ${e^{i{k_y}y}}$ can be assumed with $k_y$ conserved during transmission. Here $A_n^i$ and $B_n^i$ are the probability amplitudes of the incoming waves from the left and right, respectively, while $A_n^o$ and $B_n^o$ are those of the outgoing waves. $a_m$ and $b_m$ are constant coefficients that can be determined by boundary conditions. ${E_n} = {E} + n\hbar \omega $ are the $n$-th order Floquet energies and ${k_{xn}} = \sqrt {{{\left( {{{{E_n}} \mathord{\left/
 {\vphantom {{{E_n}} {\hbar {v_F}}}} \right.
 \kern-\nulldelimiterspace} {\hbar {v_F}}}} \right)}^2} - k_y^2} $ are the corresponding wave vectors. The latter are imaginary for evanescent modes. ${q_m} = \sqrt {{{\left[ {{{\left( {{E_m} + {V_0}} \right)} \mathord{\left/
 {\vphantom {{\left( {{E_m} - {V_0}} \right)} {\hbar {v_F}}}} \right.
 \kern-\nulldelimiterspace} {\hbar {v_F}}}} \right]}^2} - k_y^2} $ and $J_n (x)$ are the $n$-th order first kind Bessel functions. $s_n ={\rm{sign}} (E_n)$, $s'_m={\rm{sign}} (E_m+V_0)$, ${\phi _n} = {\tan ^{ - 1}}\left( {{{{k_y}} \mathord{\left/
 {\vphantom {{{k_y}} {{k_{xn}}}}} \right.
 \kern-\nulldelimiterspace} {{k_{xn}}}}} \right)$, and ${\theta _m} = {\tan ^{ - 1}}\left( {{{{k_y}} \mathord{\left/
 {\vphantom {{{k_y}} {{q_m}}}} \right.
 \kern-\nulldelimiterspace} {{q_m}}}} \right)$.

By continuity of the spinor wave function at the two boundaries $x=0$ and $x=L$, we can obtain the matrix equation (see the Appendix):
\begin{equation}
\left( {\begin{array}{*{20}{c}}
   {A_n^o}  \\
   {B_n^o}  \\
\end{array}} \right) = \sum\limits_m {{{\bf{S}}_{nm}}\left( {\begin{array}{*{20}{c}}
   {A_m^i}  \\
   {B_m^i}  \\
\end{array}} \right)} .
\end{equation}
Considering the real current flux, the Floquet scattering matrix follows as
\begin{equation}
{\bf{s}}\left( {{E_n},{E_m}} \right) = \sqrt {\frac{{{\mathop{\rm Re}\nolimits} \left( {{k_{xn}}} \right)}}{{{\mathop{\rm Re}\nolimits} \left( {{k_{xm}}} \right)}}} {{\bf{S}}_{nm}} = \left( {\begin{array}{*{20}{c}}
   {{r_{nm}}} & {t{'_{nm}}}  \\
   {{t_{nm}}} & {r{'_{nm}}}  \\
\end{array}} \right),
\label{DefineScatteringMatrix}
\end{equation}
with $r_{nm}$ and $t_{nm}$ the reflection and transmission amplitudes from the $m$th Floquet channel to the $n$th Floquet channel, respectively. ${r'}_{nm}$ and ${t'}_{nm}$ are the corresponding backward amplitudes. The scattering matrix element vanishes for evanescent modes with imaginary incoming or outgoing wave vector.

From the scattering matrix $\bf{ s}$, the total transmission probability $T_F$ can be defined as
\begin{equation}
{T_F} = \sum\limits_{n =  - \infty }^{ + \infty } {{{\left| {{t_{0n}}} \right|}^2}} =\sum\limits_{n = - \infty}^{+ \infty} {{{\left| {{s_{RL}}\left( {{E_F},{E_n}} \right)} \right|}^2}},
\end{equation}
with $s_{RL}$ the relative matrix element of $\bf{ s}$. The minimum number of sidebands $N$ that need to be included is determined by the strength of the oscillation from $N > V_1/ (\hbar \omega)$. In our numerical treatment to the monolayer graphene, sideband cutoff $N=2$ is taken into account, which is justified by ${V_1} \ll \hbar \omega $.

Numerical results of $T_F$ at certain parameters are shown in panel (a) of Fig. 5. Standard Fano resonance pattern is obvious when the $-1$-st order Floquet sideband coincides with the shallowest quasibound state within the quantum well. Therefore, the Fano resonance occurring position $E_{\rm{Fano}}$ is determined by the quasibound energy $E_b$ and the Floquet sideband interval $\hbar \omega$. During transmission, the transverse motion is conserved and its energy is carried from one of the reservoirs, through the quasibound level, and into the other reservoir. The Fano resonance occurs when
\begin{equation}
E_{\rm{Fano}}- \hbar \omega = E_b,
\end{equation}
which is equivalent to
${k_{x - 1}} = {k_{bx}}$ with $k_{bx}$ the $x$-component wave vector of the quasibound level within the well. If we increase the driving frequency, more energy is injected into the transport process and the sideband interval $\hbar \omega$ is increased. Then it is possible that deeper quasibound states can be activated into the transport process. Numerically we consider $\hbar \omega =20.5$ meV. By the solid red squares in Fig. 4, the Fano resonance occurring Floquet level energy $E_{\rm{Fano}} - \hbar \omega $ is marked. It covers all the quasibound levels from top of the well into $1 \hbar \omega$ deep except too weak Fano resonances.

\subsection{Pumped Shot Noise}

The Fano resonance in the transmission probabilities can be observed in the pumped current or shot noise.
In the graphene monolayer driven by single oscillating potential barrier, spatial and time-reversal symmetry secures vanishing pumped current.
The nonadiabatic pumped shot noise $S_{\alpha \beta }$ can be investigated by Eq. (\ref{Noise}) as well.
Current flux conservation secures that $S_{LL}=S_{RR}=-S_{LR}=-S_{RL}$. We consider one of the four and label $S_{LL}$ as $S$. To magnify the resonance spectrum, we also consider the derivatives of the noise over the Fermi energy with
${S_d} = {{\partial S} \mathord{\left/
 {\vphantom {{\partial S} {\partial {E_F}}}} \right.
 \kern-\nulldelimiterspace} {\partial {E_F}}}$.

Numerical results of $S$ and $S_d$ are shown in Fig. 5. At the Fano resonance Fermi energy, an inflection can be seen in the pumped shot noise, which originates from transport of all energy channels below the Fermi energy. The influence of the Fano resonance is thus weakened. Sharp Fano resonance reappears in the differentiate pumped shot noise.

Shot noise is a result of current fluctuations. Properties of the charge carriers, the conducting materials, and the potential configurations are imprinted in the shot noise, sometimes even more prominently than the conductance. In our consideration, no time-averaged current is present and the pumped shot noise especially its derivatives prominently demonstrate the Fano resonance in transmission. The complex eigenenergy profiles of graphene are reflected in the noise spectrum of the simple single-well device.

\section{Conclusions}

In conclusion, Fano resonance is found in the nonadiabatic pumped shot noise driven by a time-dependent quantum well in the 2DEG and graphene. The main results including three points. Firstly, when nonadiabatic quantum pumping is considered in two dimension with transverse motion entering, Fano resonance occurs at transport wave vector matching Fermi energies between one of the Floquet sidebands and one of the quasibound levels of the scatterer. In 2DEG, the incident electron emits an energy quantum of $\hbar \omega$, releases the transverse motion energy ${{{\hbar ^2}k_y^2} \mathord{\left/
 {\vphantom {{{\hbar ^2}k_y^2} {2{m^*}}}} \right.
 \kern-\nulldelimiterspace} {2{m^*}}}$, enters the quasibound state, supplies the transverse motion energy for the bound state ${{{\hbar ^2}k_y^2} \mathord{\left/
 {\vphantom {{{\hbar ^2}k_y^2} {2{m^*}}}} \right.
 \kern-\nulldelimiterspace} {2{m^*}}}$, and bounces back to the $E_F$ channel. This path interferes with direct tunneling giving rise to a Fano resonance at ${E_{{\rm{Fano}}}} - \hbar \omega  - {{{\hbar ^2}k_y^2} \mathord{\left/
 {\vphantom {{{\hbar ^2}k_y^2} {{m^*}}}} \right.
 \kern-\nulldelimiterspace} {{m^*}}} = {E_b}$. In graphene, the transverse energy is carried by the electron or hole during transmission. The Fano resonance occurs at $E_{\rm{Fano}}- \hbar \omega = E_b$ or equivalently ${k_{x - 1}} = {k_{bx}}$. Secondly, the complex quasibound level dispersion of graphene is imprinted by the Fano resonance in the total Floquet transmission spectrum. Thirdly, the nonadiabatic pumped shot noise measuring current fluctuations is a result of virtual transport process within a driving cycle. It can be considerably large even when the pumped current vanishes due to spatial and time-reversal symmetry. Prominent Fano resonance can be observed in the differential pumped shot noise. The complex eigenenergy configuration of graphene is reflected by the Fano resonance in the noise spectrum.

\section{Acknowledgements}

This project was supported by the National Natural Science
Foundation of China (No. 11004063 and No. 11174168), the Fundamental Research
Funds for the Central Universities, SCUT (No. 2014ZG0044), and the National Basic Research Program of China (No. 2011CB606405).

\section{Appendix: Derivation of the Floquet Scattering Matrix in Graphene }

In this Appendix, we show the detailed derivation of the Floquet scattering matrix in graphene from the spinor wave function continuity relations using the matrix format. Continuity equations of the spinor wave functions defined in Eqs. (\ref{WaveFunction1}) and (\ref{WaveFunction2}) are
\begin{equation}
A_n^i{e^{ - i{k_{xn}}L/2}} + A_n^o{e^{i{k_{xn}}L/2}} = \sum\limits_{m =  - \infty }^{ + \infty } {\left( {{a_m}{e^{ - i{q_m}L/2}} + {b_m}{e^{i{q_m}L/2}}} \right){J_{n - m}}\left( {\frac{{{V_1}}}{{\hbar \omega }}} \right)} ,
\end{equation}
\begin{equation}
\begin{array}{l}
 A_n^i{s_n}{e^{ - i{k_{xn}}L/2 + i{\phi _n}}} - A_n^o{s_n}{e^{i{k_{xn}}L/2 - i{\phi _n}}} \\
  = \sum\limits_{m =  - \infty }^{ + \infty } {\left( {s{'_m}{a_m}{e^{ - i{q_m}L/2 + i{\theta _m}}} - s{'_m}{b_m}{e^{i{q_m}L/2 - i{\theta _m}}}} \right){J_{n - m}}\left( {\frac{{{V_1}}}{{\hbar \omega }}} \right)} , \\
 \end{array}
\end{equation}
\begin{equation}
B_n^i{e^{ - i{k_{xn}}L/2}} + B_n^o{e^{i{k_{xn}}L/2}} = \sum\limits_{m =  - \infty }^{ + \infty } {\left( {{a_m}{e^{i{q_m}L/2}} + {b_m}{e^{ - i{q_m}L/2}}} \right){J_{n - m}}\left( {\frac{{{V_1}}}{{\hbar \omega }}} \right)} ,
\end{equation}
\begin{equation}
\begin{array}{l}
  - B_n^i{s_n}{e^{ - i{k_{xn}}L/2 - i{\phi _n}}} + B_n^o{s_n}{e^{i{k_{xn}}L/2 + i{\phi _n}}} \\
  = \sum\limits_{m =  - \infty }^{ + \infty } {\left( {s{'_m}{a_m}{e^{i{q_m}L/2 + i{\theta _m}}} - s{'_m}{b_m}{e^{ - i{q_m}L/2 - i{\theta _m}}}} \right){J_{n - m}}\left( {\frac{{{V_1}}}{{\hbar \omega }}} \right)} . \\
 \end{array}
\end{equation}
We define relative matrices with their elements
\begin{equation}
{\left( {{\bf{M}}_{sa}^ \pm } \right)_{nm}} = \left[ \begin{array}{l}
 \left( {s{'_m}{e^{i{\theta _m}}} + {s_n}{e^{ - i{\phi _n}}}} \right){e^{ - i{q_m}L/2}} \\
  \pm \left( {{s_n}{e^{i{\phi _n}}} - s{'_m}{e^{i{\theta _m}}}} \right){e^{i{q_m}L/2}} \\
 \end{array} \right]{J_{n - m}}\left( {\frac{{{V_1}}}{{\hbar \omega }}} \right),
\end{equation}
\begin{equation}
{\left( {{\bf{M}}_{sb}^ \pm } \right)_{nm}} = \left[ \begin{array}{l}
 \left( {{s_n}{e^{ - i{\phi _n}}} - s{'_m}{e^{ - i{\theta _m}}}} \right){e^{i{q_m}L/2}} \\
  \pm \left( {{s_n}{e^{i{\phi _n}}} + s{'_m}{e^{ - i{\theta _m}}}} \right){e^{ - i{q_m}L/2}} \\
 \end{array} \right]{J_{n - m}}\left( {\frac{{{V_1}}}{{\hbar \omega }}} \right),
\end{equation}
\begin{equation}
{\left( {{{\bf{M}}_r}} \right)_{nm}} = 2\cos \left( {{\phi _n}} \right){s_n}{e^{ - i{k_{xn}}L/2}}{\delta _{n,m}},
\end{equation}
\begin{equation}
{\left( {{{\bf{M}}_i}} \right)_{nm}} = {e^{ - i{k_{xn}}L}}{\delta _{n,m}},
\end{equation}
\begin{equation}
{\left( {{\bf{M}}_c^ \pm } \right)_{nm}} = {e^{ - \frac{{i\left( {{k_{xn}} \pm {q_m}} \right)L}}{2}}}{J_{n - m}}\left( {\frac{{{V_1}}}{{\hbar \omega }}} \right).
\end{equation}
After some algebra, it could be obtained that
\begin{equation}
\left\{ \begin{array}{l}
 {{\bf{A}}^o} = {\bf{M}}_c^ + {\bf{a}} + {\bf{M}}_c^ - {\bf{b}} - {{\bf{M}}_i}{{\bf{A}}^i}, \\
 {{\bf{B}}^o} = {\bf{M}}_c^ - {\bf{a}} + {\bf{M}}_c^ + {\bf{b}} - {{\bf{M}}_i}{{\bf{B}}^i}, \\
 \end{array} \right.
\end{equation}
\begin{equation}
\left( {\begin{array}{*{20}{c}}
   {\bf{a}}  \\
   {\bf{b}}  \\
\end{array}} \right) = \left( {\begin{array}{*{20}{c}}
   {{{\bf{a}}_A}} & {{{\bf{a}}_B}}  \\
   {{{\bf{b}}_A}} & {{{\bf{b}}_B}}  \\
\end{array}} \right)\left( {\begin{array}{*{20}{c}}
   {{{\bf{A}}^i}}  \\
   {{{\bf{B}}^i}}  \\
\end{array}} \right),
\end{equation}
with
\begin{equation}
\begin{array}{l}
 {{\bf{a}}_A} = {\left[ {{{\left( {{\bf{M}}_{sb}^ + } \right)}^{ - 1}}{\bf{M}}_{sa}^ +  - {{\left( {{\bf{M}}_{sb}^ - } \right)}^{ - 1}}{\bf{M}}_{sa}^ - } \right]^{ - 1}}\left[ {{{\left( {{\bf{M}}_{sb}^ + } \right)}^{ - 1}} - {{\left( {{\bf{M}}_{sb}^ - } \right)}^{ - 1}}} \right]{{\bf{M}}_r}, \\
 {{\bf{a}}_B} = {\left[ {{{\left( {{\bf{M}}_{sb}^ + } \right)}^{ - 1}}{\bf{M}}_{sa}^ +  - {{\left( {{\bf{M}}_{sb}^ - } \right)}^{ - 1}}{\bf{M}}_{sa}^ - } \right]^{ - 1}}\left[ {{{\left( {{\bf{M}}_{sb}^ + } \right)}^{ - 1}} + {{\left( {{\bf{M}}_{sb}^ - } \right)}^{ - 1}}} \right]{{\bf{M}}_r}, \\
 \end{array}
\end{equation}
and
\begin{equation}
\begin{array}{l}
 {{\bf{b}}_A} = {\left[ {{{\left( {{\bf{M}}_{sa}^ + } \right)}^{ - 1}}{\bf{M}}_{sb}^ +  - {{\left( {{\bf{M}}_{sa}^ - } \right)}^{ - 1}}{\bf{M}}_{sb}^ - } \right]^{ - 1}}\left[ {{{\left( {{\bf{M}}_{sa}^ + } \right)}^{ - 1}} - {{\left( {{\bf{M}}_{sa}^ - } \right)}^{ - 1}}} \right]{{\bf{M}}_r}, \\
 {{\bf{b}}_B} = {\left[ {{{\left( {{\bf{M}}_{sa}^ + } \right)}^{ - 1}}{\bf{M}}_{sb}^ +  - {{\left( {{\bf{M}}_{sa}^ - } \right)}^{ - 1}}{\bf{M}}_{sb}^ - } \right]^{ - 1}}\left[ {{{\left( {{\bf{M}}_{sa}^ + } \right)}^{ - 1}} + {{\left( {{\bf{M}}_{sa}^ - } \right)}^{ - 1}}} \right]{{\bf{M}}_r}. \\
 \end{array}
\end{equation}
The scattering matrix without flux normalization follows as
\begin{equation}
\left( {\begin{array}{*{20}{c}}
   {{{\bf{A}}^o}}  \\
   {{{\bf{B}}^o}}  \\
\end{array}} \right) = \left( {\begin{array}{*{20}{c}}
   {{{\bf{M}}_{AA}}} & {{{\bf{M}}_{AB}}}  \\
   {{{\bf{M}}_{BA}}} & {{{\bf{M}}_{BB}}}  \\
\end{array}} \right)\left( {\begin{array}{*{20}{c}}
   {{{\bf{A}}^i}}  \\
   {{{\bf{B}}^i}}  \\
\end{array}} \right) \equiv {\bf{S}}\left( {\begin{array}{*{20}{c}}
   {{{\bf{A}}^i}}  \\
   {{{\bf{B}}^i}}  \\
\end{array}} \right),
\end{equation}
with
\begin{equation}
\begin{array}{l}
 {{\bf{M}}_{AA}} = {\bf{M}}_c^ + {{\bf{a}}_A} + {\bf{M}}_c^ - {{\bf{b}}_A} - {{\bf{M}}_i}, \\
 {{\bf{M}}_{AB}} = {\bf{M}}_c^ + {{\bf{a}}_B} + {\bf{M}}_c^ - {{\bf{b}}_B}, \\
 {{\bf{M}}_{BA}} = {\bf{M}}_c^ - {{\bf{a}}_A} + {\bf{M}}_c^ + {{\bf{b}}_A}, \\
 {{\bf{M}}_{BB}} = {\bf{M}}_c^ - {{\bf{a}}_B} + {\bf{M}}_c^ + {{\bf{b}}_B} - {{\bf{M}}_i}. \\
 \end{array}
\end{equation}

\clearpage

\begin{figure}[h]
\includegraphics[height=12cm, width=14cm]{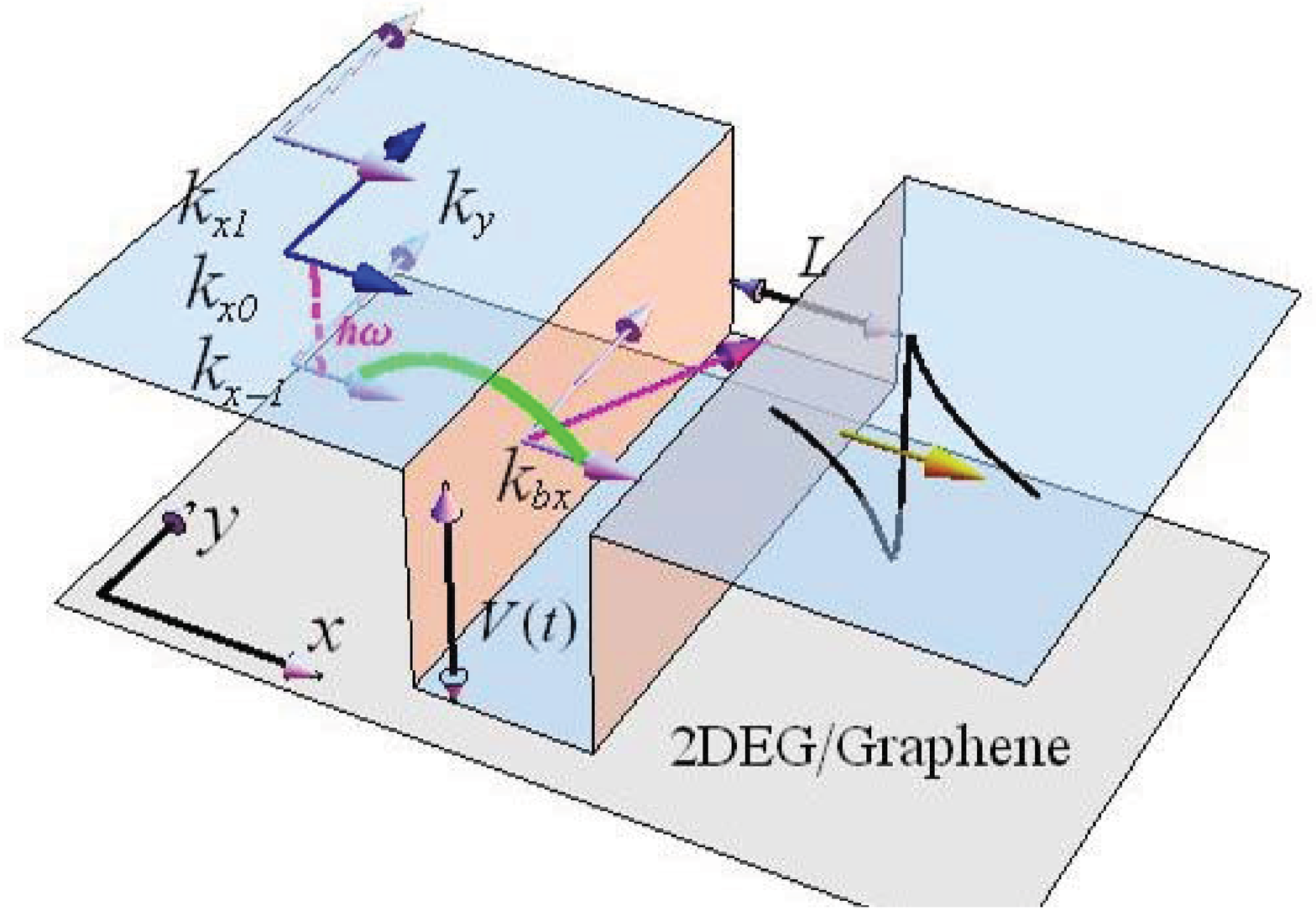}
\caption{Potential profile on the 2DEG/graphene spanned in the $x$-$y$ plane. The $y$-direction is infinite. Quasiparticles transport in the $x$ direction. A time-dependent single-well potential $V(t)=-V_0+V_1 \cos (\omega t)$ is applied with width $L$. As a dynamic effect, Floquet sidebands are formed with energy spacing $\hbar \omega$. During transmission $k_y$ is preserved. When the wave vector of one of the Floquet sidebands in the transport direction matches that of the quasibound state in the well, Fano resonance occurs in the transmission spectrum. }
\end{figure}

\begin{figure}[h]
\includegraphics[height=10cm, width=14cm]{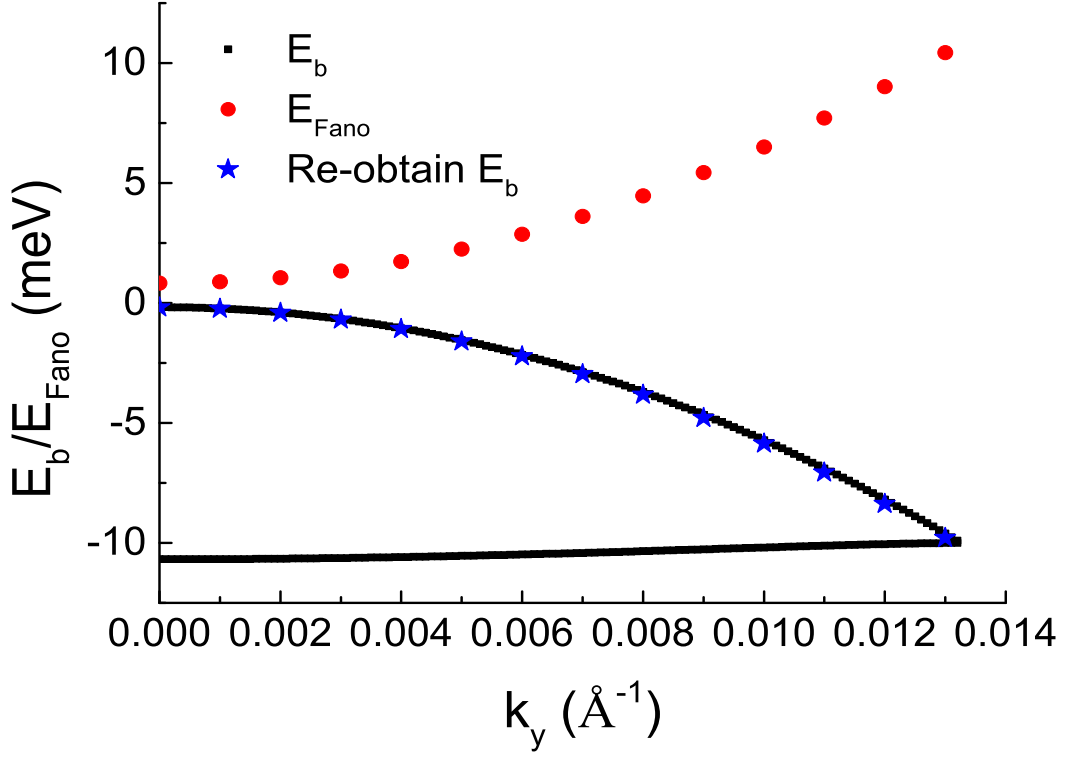}
\caption{ Solid black squares and red circles are dispersion of the quasibound states $E_b$ of the static quantum well with depth $V_0$ and width $L$ in 2DEG and that of the Fano resonance Fermi energy $E_{\rm{Fano}}$ driven by time-dependent oscillation of the quantum well, respectively. The blue asterisks are the quasibound energy reobtained from the Fano resonance energy by Eq. (\ref{FanoEb}). Numerical parameters are\cite{LiPRB1999, DaiEPJB2014} $\hbar \omega =$1 meV, $L=10$ $\rm{\AA}$, $V_0=$20 meV, $V_1=$5 meV. Floquet sideband cutoff $N=$5.  }
\end{figure}

\begin{figure}[h]
\includegraphics[height=11cm, width=15cm]{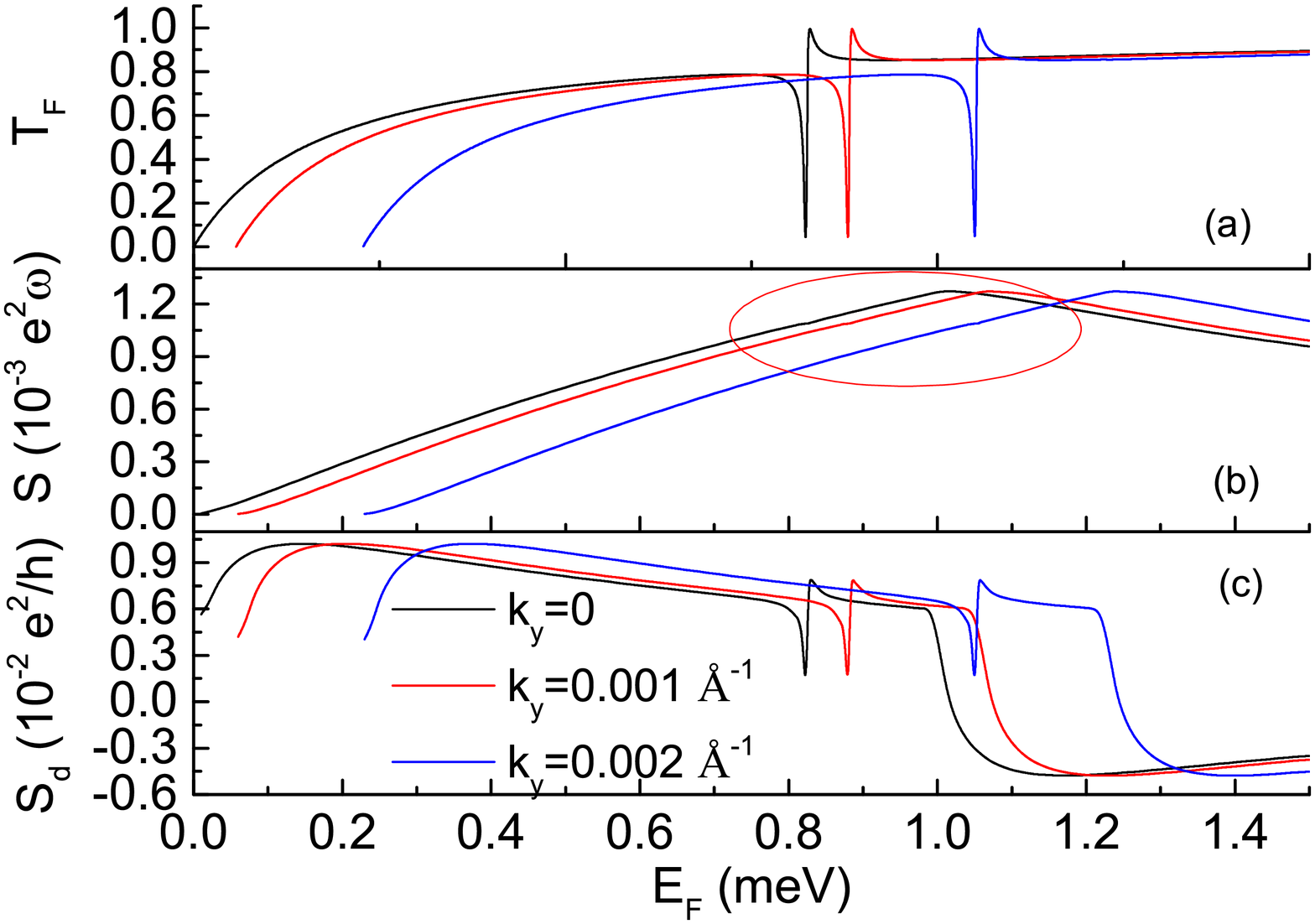}
\caption{  (a) Total Floquet transmission probability $T_F$, (b) pumped shot noise $S$, and (c) differential pumped shot noise $S_d$ driven by an oscillating potential well as a function of $E_F$ for different $k_y$ in 2DEG. An inflection occurs in the pumped shot noise corresponding to the Fano resonance in the
transmission highlighted by the red circle in panel (b). Sharp resonance could be seen at the inflection energy in panel (c). Parameters are the same as Fig. 2. The noise units are obtained by substituting $\hbar \omega$ = 1 meV into the
energy and absorbing the additional $2 \pi $ into the data.  }
\end{figure}

\begin{figure}[h]
\includegraphics[height=10cm, width=14cm]{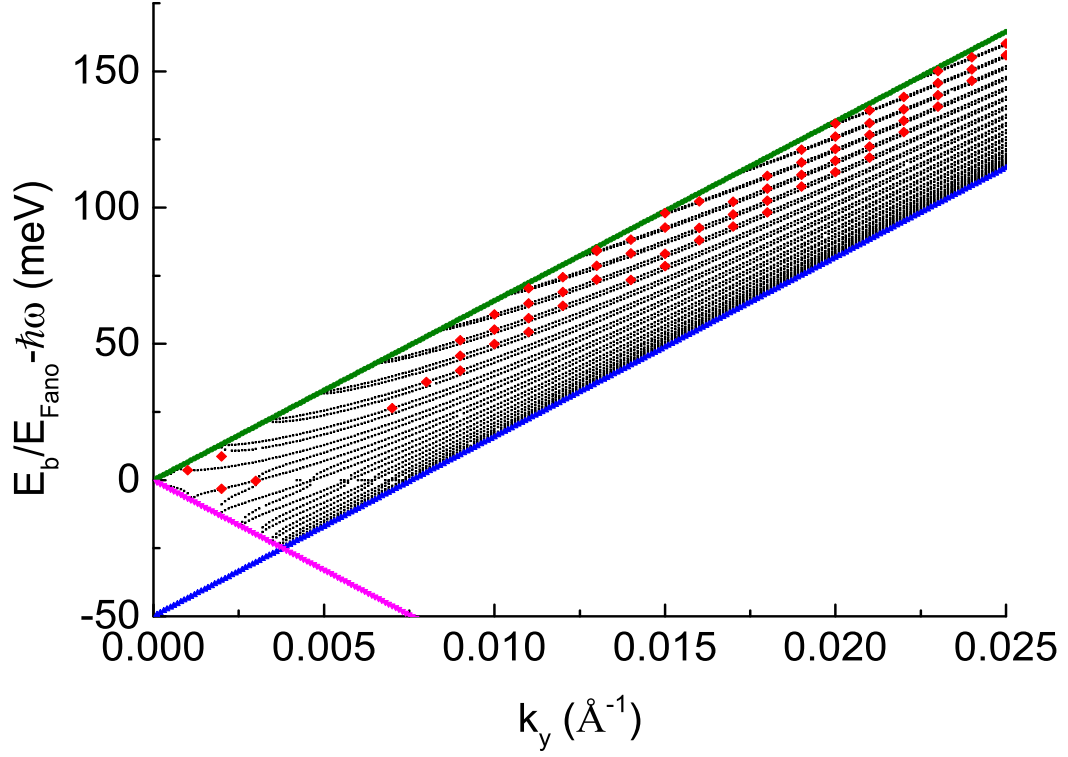}
\caption{ Small black dots are dispersion of the quasibound states $E_b$ of the static quantum well with depth $V_0$ and width $L$ in graphene. Green, blue, and pink lines are the Dirac ``light-cone" band boundaries. The red solid squares are the quasibound energy $E_b$ reobtained from the Fano resonance energy $E_{\rm{Fano}}$ by $E_{\rm{Fano}}-\hbar \omega$. Numerical parameters are $\hbar \omega =20.5$ meV, $L=3000$ $\rm{\AA}$, $V_0=50$ meV, $V_1=1$ meV. Floquet sideband cutoff $N=2$.  }
\end{figure}

\begin{figure}[h]
\includegraphics[height=11cm, width=15cm]{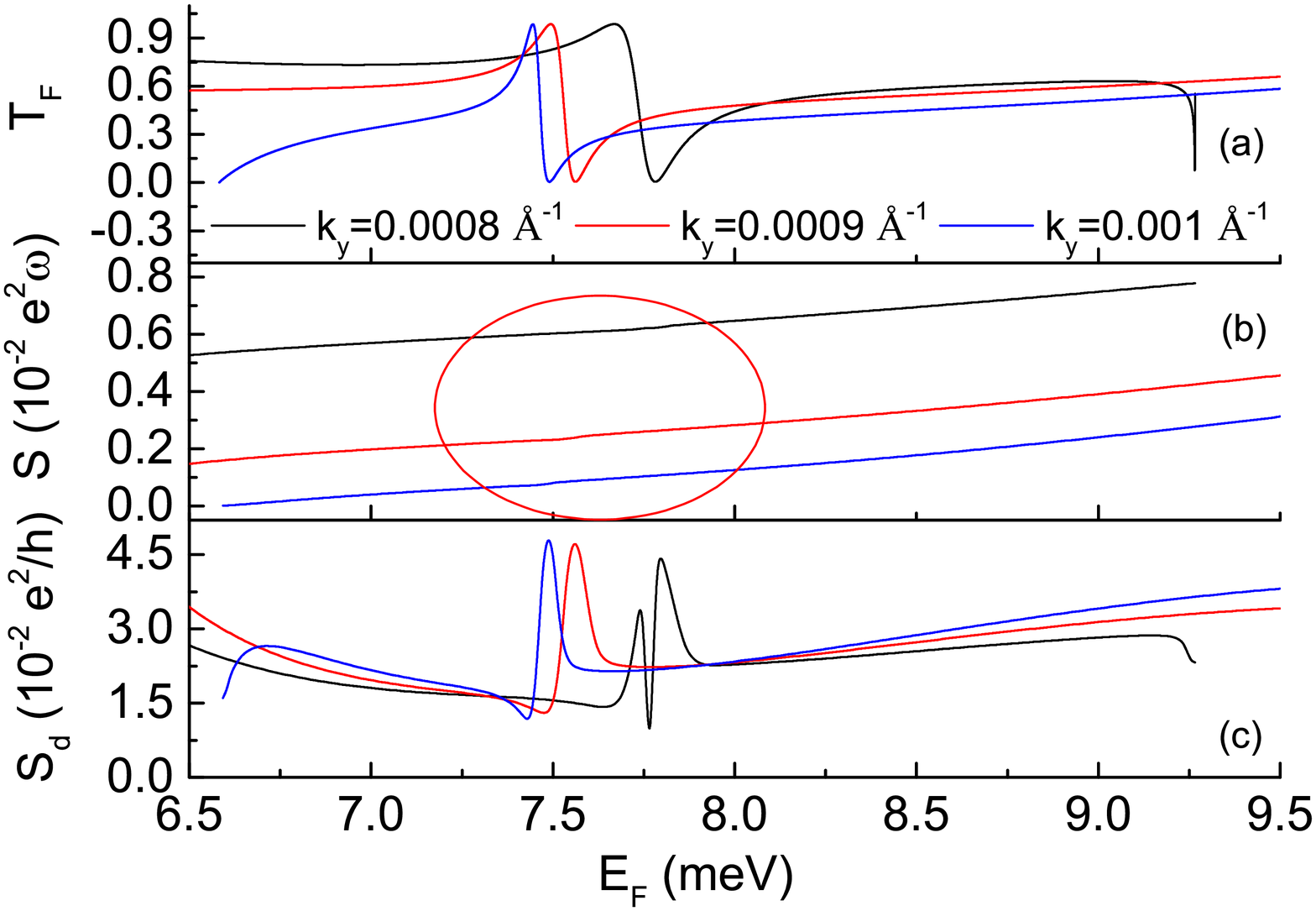}
\caption{  (a) Total Floquet transmission probability $T_F$, (b) pumped shot noise $S$, and (c) differential pumped shot noise $S_d$ driven by an oscillating potential well as a function of $E_F$ for different $k_y$ in graphene. An inflection occurs in the pumped shot noise corresponding to the Fano resonance in the
transmission highlighted by the red circle in panel (b). Sharp resonance could be seen at the inflection energy in panel (c). $\hbar \omega =4$ meV and other parameters are the same as Fig. 4. The noise units are obtained by substituting $\hbar \omega$ = 4 meV into the
energy and absorbing the additional $2 \pi $ into the data.  }
\end{figure}

\clearpage

\clearpage

\end{document}